\documentclass[prc,showpacs,superscriptaddress,twocolumn]{revtex4}

\usepackage{amsmath,bm}
\usepackage{graphicx}
\usepackage{epstopdf}
\usepackage{subfigure}
\usepackage{epsfig}
\usepackage{amsmath,amssymb,amsfonts}
\usepackage{color}
\usepackage[utf8]{inputenc}
\usepackage{verbatim}
\usepackage{array}
\usepackage{hyperref}
\graphicspath{{figures/}} % Directory in which figures are stored

\begin{document}

\renewcommand{\topfraction}{0.85}
\renewcommand{\textfraction}{0.1}
\renewcommand{\floatpagefraction}{0.75}

\title{Probing the mass effect of heavy quark jets in high-energy nuclear collisions}

\author{Sa Wang}
%\email{wangsa@ctgu.edu.cn}
\affiliation{College of Science, China Three Gorges University, Yichang 443002, China}
\affiliation{Center for Astronomy and Space Sciences and Institute of Modern Physics, China Three Gorges University, Yichang 443002, China}
\affiliation{Key Laboratory of Quark \& Lepton Physics (MOE) and Institute of Particle Physics, Central China Normal University, Wuhan 430079, China}

\author{Shuang Li}
%\email{lish@ctgu.edu.cn}
\affiliation{College of Science, China Three Gorges University, Yichang 443002, China}
\affiliation{Center for Astronomy and Space Sciences and Institute of Modern Physics, China Three Gorges University, Yichang 443002, China}

\author{Yao Li}
\affiliation{Key Laboratory of Quark \& Lepton Physics (MOE) and Institute of Particle Physics, Central China Normal University, Wuhan 430079, China}

\author{Ben-Wei Zhang}
\email{bwzhang@mail.ccnu.edu.cn}
\affiliation{Key Laboratory of Quark \& Lepton Physics (MOE) and Institute of Particle Physics, Central China Normal University, Wuhan 430079, China}

\author{Enke Wang}
%\email{wangek@scnu.edu.cn}
\affiliation{Guangdong-Hong Kong Joint Laboratory of Quantum Matter, Guangdong Provincial Key Laboratory of Nuclear Science, Southern Nuclear Science Computing Center, South China Normal University, Guangzhou 510006, China}

\date{\today}

%%%%%%%%%%%%%%%%%%%%%%%%%%%%%%%%%%%%%%%%%%%%%%%%%%%%%%%%%%%%%%%%%%%%%
\begin{abstract}
The production of heavy quark (HQ) jets provides a new arena to address the mass effect of jet quenching in heavy-ion physics. This paper presents a theoretical study of HQ jet yield suppression in Pb+Pb collisions at the LHC and focuses on the energy loss of HQ jets produced by different mechanisms. The p+p baseline is carried out by the SHERPA generator, and the jet-medium interactions are described by the SHELL transport model, which considers the elastic and inelastic partonic energy loss in the quark-gluon plasma (QGP). In p+p collisions, our numerical results indicate that the HQ jets from gluon splitting ($g \rightarrow Q$-jet) give the dominant contribution at high $p_T$, and it shows more dispersive structures than the HQ-initiated one ($Q \rightarrow Q$-jet). In nucleus-nucleus collisions, our calculations are consistent with the inclusive and b-jet $R_{AA}$ recently measured by the ATLAS collaboration, which suggests a remarkable manifestation of the mass effect of jet energy loss. As a result of the dispersive substructure, the $g \rightarrow Q$-jet will lose more energy than the $Q \rightarrow Q$-jet in the QGP. Due to the significant contribution of $g \rightarrow c$-jet, the $R_{AA}$ of c-jet will be comparable or even smaller than that of inclusive jet. To experimentally distinguish the $g \rightarrow Q$-jet and $Q \rightarrow Q$-jet, we propose the event selection strategies based on their topological features and test the performances. By isolating the $c \rightarrow c$-jet and $b \rightarrow b$-jet, the jets initiated by heavy quarks, we predict that the order of their $R_{AA}$ are in line with the mass hierarchy of energy loss. Future measurements on the $R_{AA}$ of $Q \rightarrow Q$-jet and $g \rightarrow Q$-jet will provide a unique chance to test the flavor/mass dependence of energy loss at the jet level.
\end{abstract}

\pacs{25.75.Ld, 25.75.Gz, 24.10.Nz}
\maketitle

%%%introduction%%%
%%%%%%%%%%%%%%%%%%%%%%%%%%%%%%%%%%%%%%%%%%%%%%%%%%%%%%%%%%%%%%%%%%%%%

\section{Introduction}
\label{sec:introduction}

High-energy nuclear collisions at the Relativistic Heavy Ion Collider (RHIC) and the Large Hadron Collider (LHC) provide an excellent arena to unravel the mysteries of the quark-gluon plasma (QGP), a new state of nuclear matter formed at extremely high temperature and density. The ``jet quenching" effect, energy attenuation of fast partons due to their strong interactions with the constituents of the QGP medium, aroused  physicists' great interest and has been extensively studied ~\cite{Gyulassy:2003mc, Gyulassy:1990ye, Qin:2015srf, Vitev:2008rz, Proceedings:2007ctk, Vitev:2009rd, Casalderrey-Solana:2014bpa, Gyulassy:1993hr,
Wang:2001ifa, Vitev:2008vk, He:2020iow, Chen:2022kic, Zhao:2021vmu, Yang:2023dwc, JETSCAPE:2022jer, Luo:2023nsi, Xie:2024xbn, Zhang:2021xib, Kang:2023ycg, Kang:2023qxb, Chen:2024cgx}. Investigations of the jet quenching phenomenon deepen our understanding of the quantum chromodynamics (QCD) under extreme conditions and reveal the properties of the strongly-coupled nuclear matter \cite{Connors:2017ptx, Andrews:2018jcm, Cunqueiro:2021wls}.

As a result of the large mass ($M_Q\gg\Lambda_{QCD}$), heavy quarks (HQ) are powerful hard probes to explore the transport properties of the QGP~\cite{Dong:2019unq, Andronic:2015wma, Dong:2019byy, Zhao:2020jqu, Chen:2023xhd, Jiang:2022uoe}. Over the past two decades, measurements on the nuclear modification factor $R_{AA}$~\cite{Adamczyk:2014uip, Adam:2015sza, Sirunyan:2017xss, ALICE:2018lyv} and the collective flow $v_n$~\cite{Abelev:2014ipa, Adamczyk:2017xur, Acharya:2017qps, Sirunyan:2017plt} of heavy-flavor hadrons have enriched our knowledge of the energy loss mechanisms and hadronization patterns of heavy quarks in high-energy heavy-ion collisions. Due to the ``dead-cone'' effect \cite{Dokshitzer:2001zm}, heavy quarks will lose less energy than the massless light partons \cite{Zhang:2003wk, Armesto:2003jh, Djordjevic:2003qk}. By comparing the $R_{AA}$ of heavy-flavor hadrons \cite{ALICE:2015nvt, CMS:2016mah, CMS:2017uoy, ATLAS:2018hqe} as well as their decayed leptons \cite{STAR:2021uzu, PHENIX:2022wim, ATLAS:2021xtw} with that of light-flavor one, some pieces of evidence of the mass effect have been addressed \cite{Xing:2019xae, Zhang:2022rby, Wang:2023udp}.

Beyond that, the HQ jets, defined as jets containing heavy-flavor quarks/hadrons \cite{LHCb:2015tna, CMS:2012feb}, are also excellent tools to capture the mass effect of energy loss at jet level \cite{Chatrchyan:2013exa, ATLAS:2022agz, Sirunyan:2018jju, Huang:2013vaa, Li:2018xuv, Kang:2018wrs, Dai:2018mhw, Li:2024uzk, Li:2024pfi}. The richer inner structure of HQ jets compared to single particles provides a unique opportunity to explore the exquisite interaction mechanisms between the hard parton and the medium. Meanwhile, investigation of the production mechanisms and substructures of HQ jets has attracted much attention on both the experimental~\cite{Sirunyan:2019dow, ALICE:2019cbr, ALICE:2021aqk, ATLAS:2021agf, Vertesi:2021brz, CMS:2020geg, ALICE:2022phr, ATLAS:2018zhf} and the theoretical sides~\cite{Goncalves:2015prv, Ilten:2017rbd, Li:2017wwc, Li:2021gjw, Wang:2019xey, Wang:2020ukj, Li:2022tcr, Wang:2023eer, Wang:2021jgm, Attems:2022otp}. The recent measurements on the jet radial profile \cite{Sirunyan:2019dow} and jet shape by CMS collaboration~\cite{CMS:2020geg} imply that HQ jets produced in different mechanisms may exhibit distinct topologies and structures~\cite{Norrbin:2000zc, Banfi:2007gu}. In this context, it would be of great interest to explore the energy loss effect of HQ jets produced by different channels and their relation to their substructures. Note that the HQ jet samples selected in the experiment include the jets initiated by heavy quarks ($Q\rightarrow Q$-jet) and a considerable contribution from the gluon splitting ($g\rightarrow Q$-jet). The former is initiated by a heavy quark created in the early stage of the QCD hard scattering, while the latter can be produced by the splitting of a high-energy gluon ($g\rightarrow Q\bar{Q}$) during the parton shower. It is still challenging to experimentally address the mass effect of jet energy loss of the HQ-initiated jet compared with the massless one. Exploring suitable selection strategies to isolate HQ jets produced by different production mechanisms is necessary, which makes it possible to directly compare the yield suppression of the HQ-initiated jets with that of the light-flavor one.

This paper studies the yield suppression of HQ jets in heavy-ion collisions to address the mass effect of jet energy loss. At first, we will estimate the fractional contributions from different production mechanisms to the HQ jet yields in p+p collisions and discuss their main characteristics in the jet substructure. In nucleus-nucleus collisions, we will systematically estimate the fractional contribution and energy loss of HQ jets from $Q\rightarrow Q$-jet and $g\rightarrow Q$-jet. We will show that a significant contribution of $g\rightarrow Q$-jet and its dispersive jet substructure will lead to a comparable $R_{AA}$ of the c-jet relative to the inclusive jet. Furthermore, to make the separation of $g \rightarrow Q$-jet and $Q \rightarrow Q$-jet accessible in experimental measurements, we will propose strategies to distinguish them based on their topological features. By comparing the yield suppression of the select $Q \rightarrow Q$-jet sample with that of the inclusive jet, we will show that the mass hierarchy of energy loss ($\Delta E_{\textit{\rm incl-jet}}>\Delta E_{\textit{\rm c-jet}}>\Delta E_{\textit{\rm b-jet}}$) at jet level holds the true.

The remainder of this paper is organized as follows. In Sec.~II, we will discuss the production mechanisms of HQ jets in p+p collisions. In Sec.~III, the theoretical frameworks of the transport model used to study the medium modification of HQ jet will be introduced. In Sec.~IV, we will show the main results and give specific discussions. At last, we will summarize this work in Sec.~IV.

%%%%%%%%%%%%%%%%%%%%%%%%%%%%%%%%%%%%%%%%%%%%%%%%%%%%%%%%%%%%%%%%%%%%%
%%%%%%%%%%%%%%%%%%%%%%%%%%%%%%%%%%%%%%%%%%%%%%%%%%%%%%%%%%%%%%%%%%%%%
%%%%%%%%%%%%%%%%%%%%%%%%%%%%%%%%%%%%%%%%%%%%%%%%%%%%%%%%%%%%%%%%%%%%%
%%%%%%%%%%%%%%%%%%%%%%%%%%%%%%%%%%%%%%%%%%%%%%%%%%%%%%%%%%%%%%%%%%%%%

\section{Heavy quark jet production in p+p collisions}
\label{sec:ppbaseline}

\begin{figure}[!t]
\begin{center}
%\vspace*{0.1in}
 \subfigure[]{\label{fig:process1}
  \epsfig{file=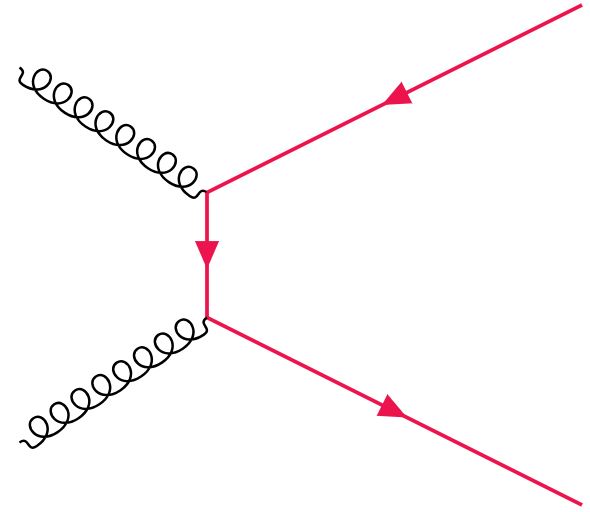, width=0.13\textwidth, clip=}}
 \subfigure[]{\label{fig:process2}
  \epsfig{file=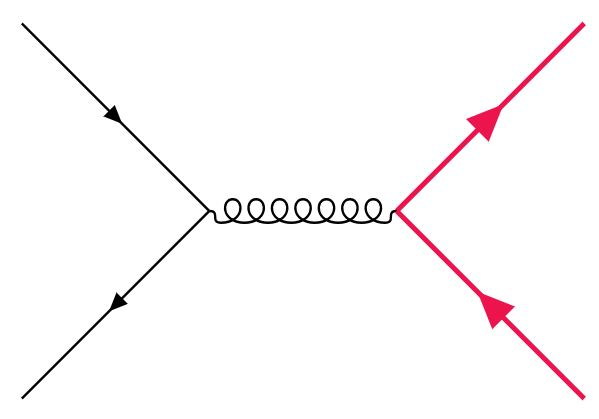, width=0.13\textwidth, clip=}}
 \subfigure[]{\label{fig:process3}
  \epsfig{file=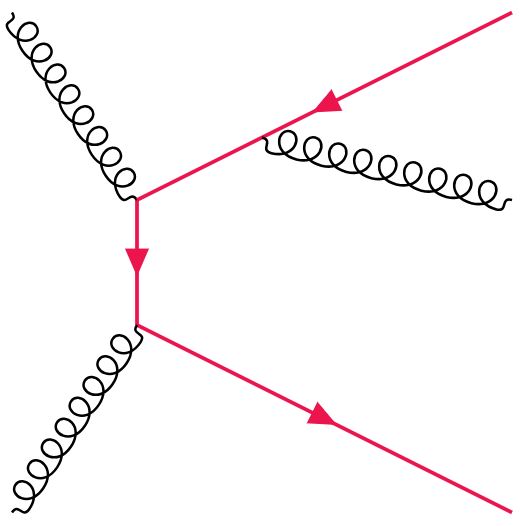, width=0.13\textwidth, clip=}}
 \subfigure[]{\label{fig:process4}
  \epsfig{file=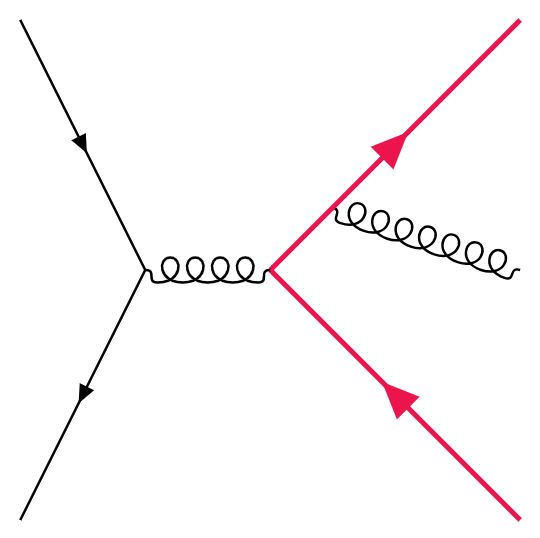, width=0.13\textwidth, clip=}}
 \subfigure[]{\label{fig:process5}
  \epsfig{file=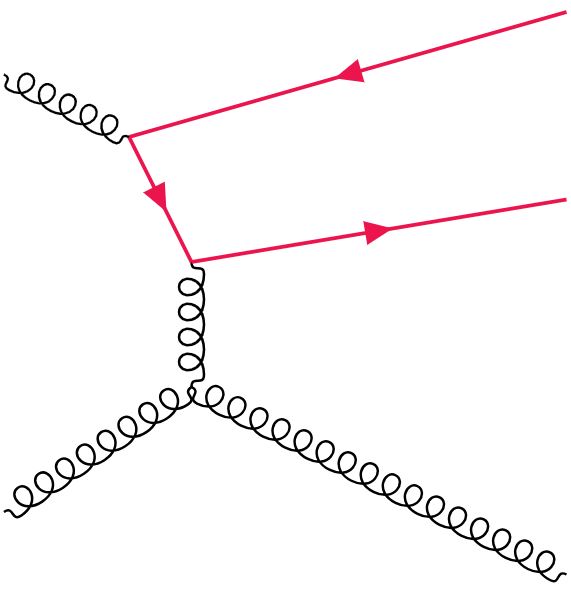, width=0.13\textwidth, clip=}}
 \subfigure[]{\label{fig:process6}
  \epsfig{file=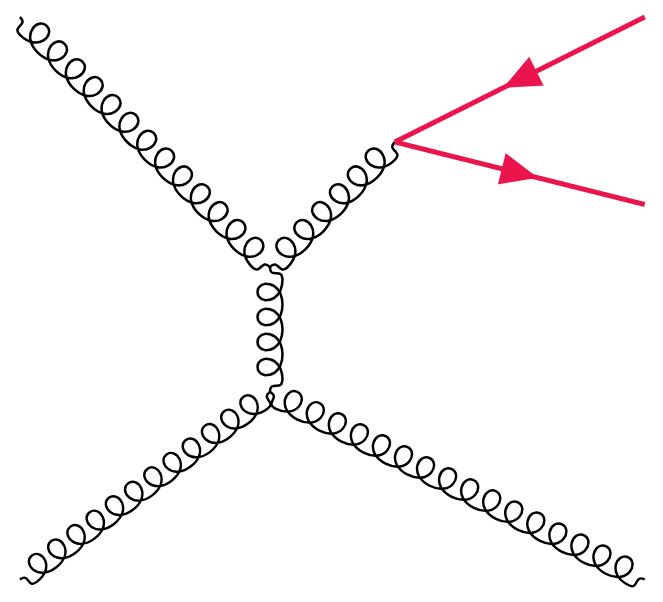, width=0.13\textwidth, clip=}}
  \caption{Typical examples of QCD Feynman diagrams contributing to the production of HQ jets. Flavor Creation: LO(a,b), NLO(c,d). Flavor Excitation: (e). Gluon Splitting:(f). The red fermion lines denote the produced heavy quarks.}
  \label{fig:process}
\end{center}
\end{figure}

\begin{figure}[!t]
\begin{center}
%\vspace*{0.1in}
\includegraphics[width=3.0in,angle=0]{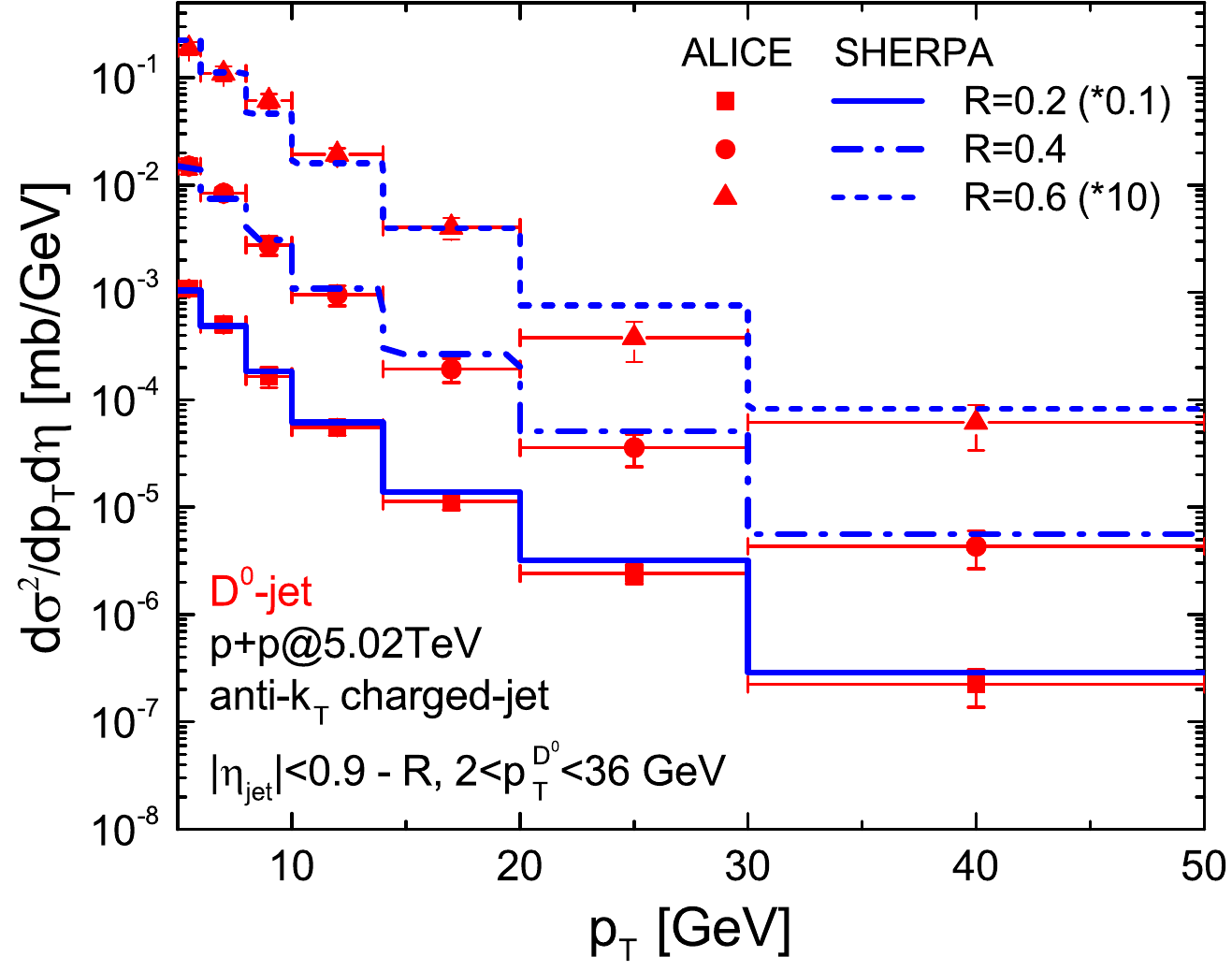}
\includegraphics[width=3.0in,angle=0]{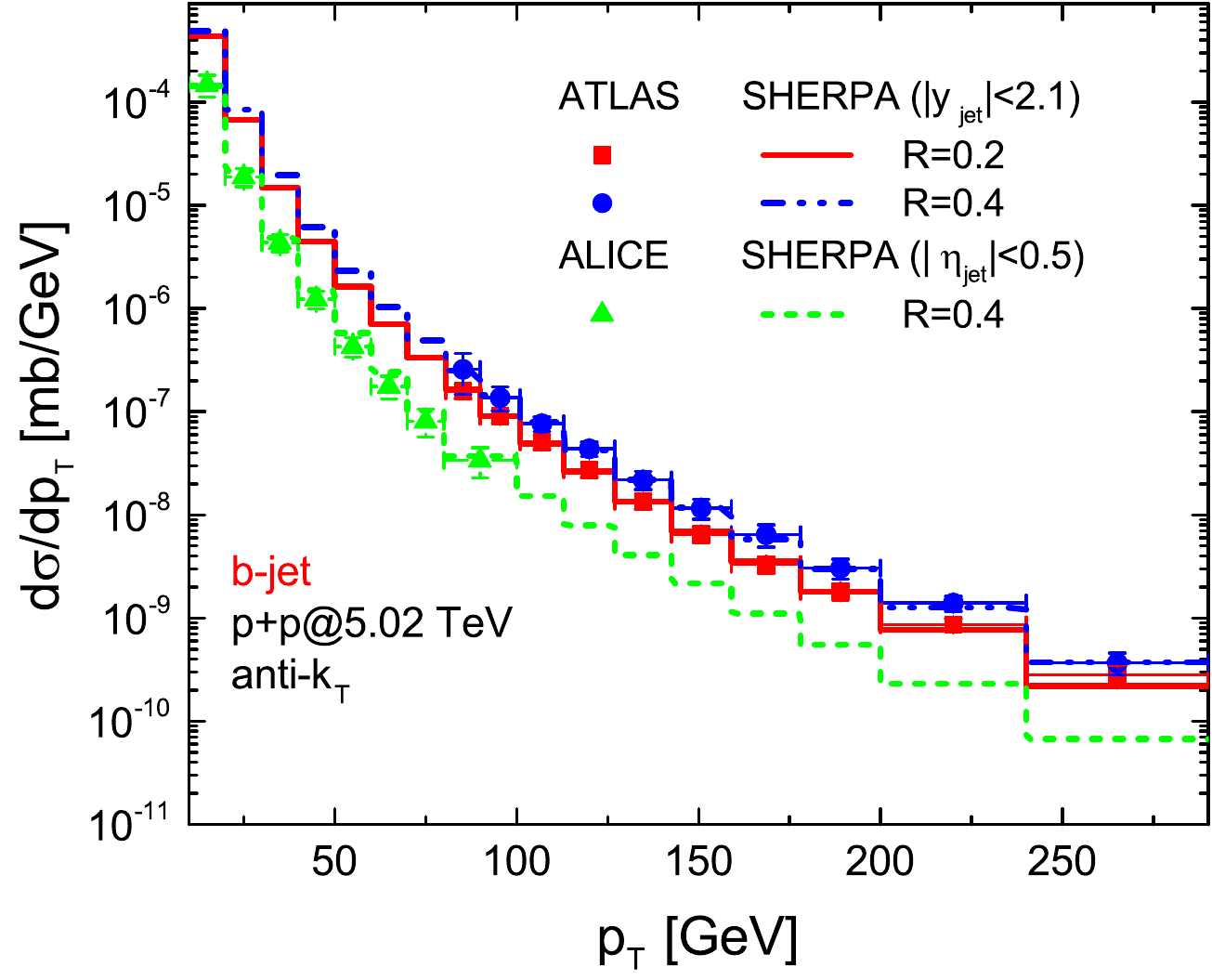}
\vspace*{0in}
\caption{(Color online) Upper panel: differential cross sections of the $D^0$ meson tagged-jet by SHERPA in p+p collisions at $\sqrt{s}=$ 5.02 TeV compared with the ALICE data~\cite{ALICE:2022mur}, at R = 0.2, 0.4, 0.6 respectively. Lower panel: differential cross section of b-jet in p+p collisions at $\sqrt{s}=$ 5.02 TeV compared with the ALICE~\cite{ALICE:2021wct} and the ATLAS~\cite{ATLAS:2022agz} data.}
\label{fig:pp-QjetXS}
\end{center}
\end{figure}

\begin{figure}[!t]
\begin{center}
\vspace*{0.1in}
\includegraphics[width=3.0in,angle=0]{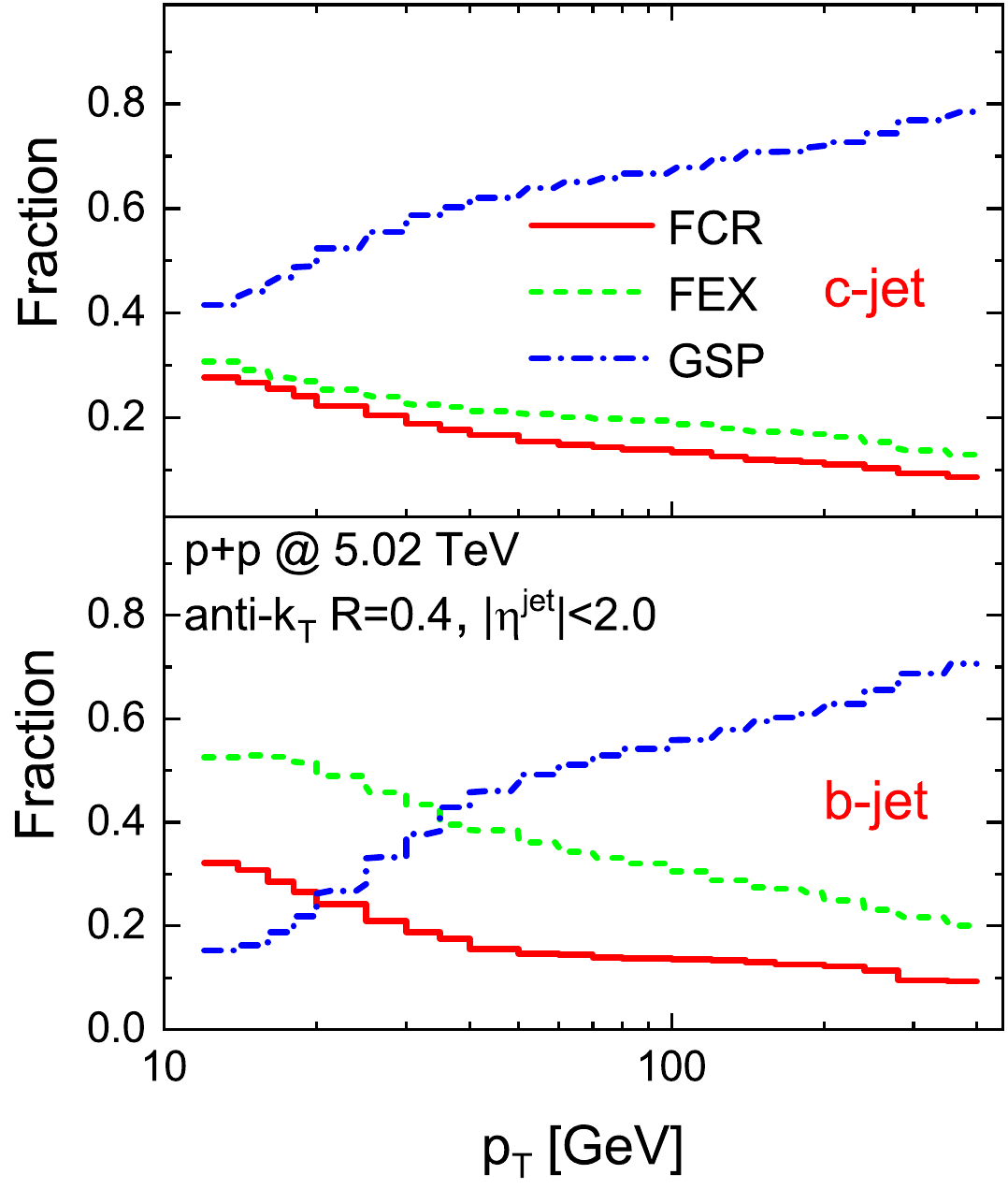}
\vspace*{-0.1in}
\caption{(Color online) The fractional contributions of the three production mechanisms to the total HQ jet differential cross section as functions of jet $p_T$ in p+p collisions at $\sqrt{s}=$ 5.02 TeV for c-jet and b-jet.}
\label{fig:pp-g-cjet}
\end{center}
\end{figure}

Production of open heavy flavors in hadron collisions has been studied by various theoretical schemes over the past decades \cite{Nason:1987xz, Nason:1989zy, Beenakker:1990maa, Aivazis:1993kh, Kniehl:2005mk, Cacciari:1998it}. We show in Fig.~\ref{fig:process} the typical QCD Feynman diagrams contributing to the production of HQ jets, which are usually categorized into three classes: the flavor creation (FCR), the flavor excitation (FEX), and the gluon splitting (GSP)~\cite{Norrbin:2000zc, Banfi:2007gu}. The first two represent the pair creation of heavy quarks in hard scattering at leading order, in which the incoming partons are light quarks or gluons, and the outcoming $Q\bar{Q}$ pairs usually emerge back-to-back in azimuth. Higher-order QCD processes of heavy quark production are also not negligible in the hadron collisions at the LHC energy. Diagrams (c) and (d) represent the FCR processes at NLO with an extra radiated gluon at the final state. The diagram (e) in Fig.~\ref{fig:process} is the typical FEX process, in which a heavy quark from the parton distribution function of one incoming proton is excited in the hard scattering by a light parton of another proton. As we can see, the most distinct feature of the FEX compared to FCR is that only one heavy quark is produced by the hard scattering. The last diagram in Fig.~\ref{fig:process} is the splitting process of a final-state hard gluon during the parton shower, and in general, the $Q\bar{Q}$ pairs are produced with the smaller opening angle in azimuth compared to the FCR. The prior clarification of the different kinematics features of these subprocesses may help us understand the HQ jets' corresponding substructures and topological features.

In this work, we employ the Monte Carlo event generator SHERPA~\cite{Gleisberg:2008ta, Sherpa:2019gpd} to compute the initial HQ jet production in p+p collisions, which matches the hard QCD processes at next-to-leading order with the vacuum parton shower effect (NLO+PS). The resummation of the parton shower based on Catani-Seymour subtraction method~\cite{Schumann:2007mg} is merged with fixed-order NLO calculation by the MC@NLO prescription~\cite{Frixione:2002ik}. The NNPDF 3.0~\cite{Ball:2008by} parton distribution functions (PDFs) have been chosen in the calculations of SHERPA. In the upper panel of Fig.~\ref{fig:pp-QjetXS}, we show the calculated differential cross section of the $D^0$ meson tagged-jet by SHERPA in p+p collisions at $\sqrt{s}=$ 5.02 TeV compared with the ALICE data~\cite{ALICE:2022mur}. The charged jets are reconstructed by using the Fastjet program~\cite{Cacciari:2011ma} with anti-$k_T$ algorithm~\cite{Cacciari:2008gp} at R = 0.2, 0.4, 0.6 and rapidity range $|\eta^{\rm jet}|< 0.9-R$. The $D^0$ mesons are required to have  2$<p_T^{D}<$ 36 GeV. In the lower panel of Fig.~\ref{fig:pp-QjetXS}, we also present the yield of b-jet in p+p collisions at $\sqrt{s}= 5.02$ TeV compared with the ATLAS and ALICE data~\cite{ATLAS:2022agz, ALICE:2021wct}. For the ATLAS measurement, the b-jets are constructed by the anti-$k_T$ jet algorithm using $R=0.2$ and $R=0.4$ within $|y_{\rm jet}|<2.1$. As for the ALICE data, the b-jet are constructed with charged particles using $R=0.4$ within $|\eta_{\rm jet}|<0.5$. We observe that the calculations by SHERPA are consistent with the experimental measurements of the HQ jet yields both for c-jet and b-jet, which establish a good baseline for subsequent studies of their nuclear modifications.

In Monte Carlo simulations, the three mechanisms mentioned in Fig.~\ref{fig:process} can be distinguished based on their topological features. The events with one and two outcoming heavy quarks in the hard processes can be categorized into FEX and FCR, respectively. In contrast, the events containing HQ jet created only in the parton shower stage are regarded as the GSP type. As shown in Fig.~\ref{fig:pp-g-cjet}, we estimate the fractional contributions of the three mechanisms to the total yield as a function of jet $p_T$ for both c-jet (upper) and b-jet (lower) in p+p collisions at $\sqrt{s}=$ 5.02 TeV. At $12<p_T<400$ GeV, the proportions of GSP increase with jet $p_T$ and eventually become the most pronounced mechanism for both c-jet and b-jet production, which is consistent with the estimation in Ref.~\cite{Banfi:2007gu}. For c-jet, at lower $p_T$ the contributions of these three mechanisms are nonnegligible. As for b-jet, one can see that at lower $p_T$, the most important (above 50 $\%$) contributions is from the FEX mechanism, while the one from the GSP is less than $20\%$. Since FCR and FEX denote the jets initiated by heavy quarks, while GSP corresponds to HQ jets initiated by high-energy gluon, to focus on their essential differences, it is convenient to categorize these three mechanisms into two subprocesses, $Q\rightarrow Q$-jet and $g\rightarrow Q$-jet, in the remainder of this paper.

\begin{figure}[t]
\begin{center}
\vspace*{0.0in}
\includegraphics[width=3.0in,angle=0]{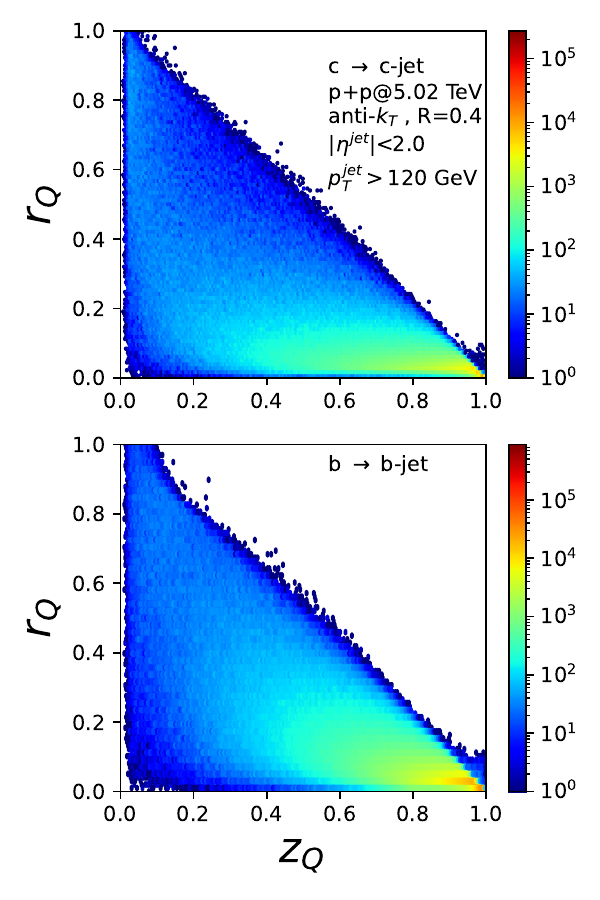}
\vspace*{0in}
\caption{(Color online) Two-dimensional ($z_Q,r_Q$) correlation diagrams of $c\rightarrow c$-jet (upper panel) and $b\rightarrow b$-jet (lower panel) in p+p collisions at $\sqrt{s}=$ 5.02 TeV.}
\vspace*{-0.2in}
\label{fig:pp-zrQ}
\end{center}
\end{figure}

\begin{figure}[t]
\begin{center}
\vspace*{0in}
\includegraphics[width=3.0in,angle=0]{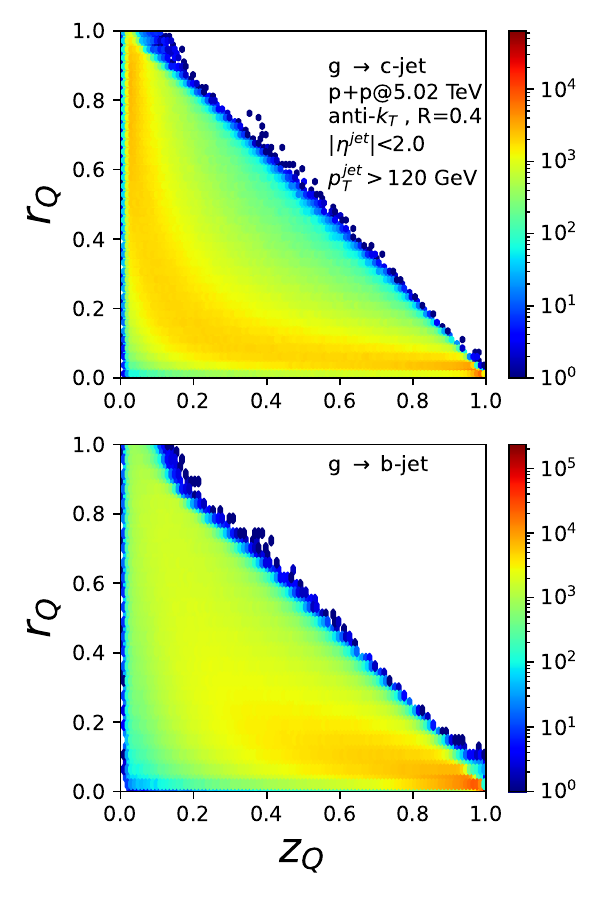}
\vspace*{0in}
\caption{(Color online) Two-dimensional ($z_Q,r_Q$) correlation diagrams of $g\rightarrow c$-jet (upper panel) and $g\rightarrow b$-jet (lower panel) in p+p collisions at $\sqrt{s}=$ 5.02 TeV.}
\vspace*{-0.2in}
\label{fig:pp-zrg}
\end{center}
\end{figure}

To show the essential differences of $Q\rightarrow Q$-jet and $g\rightarrow Q$-jet in substructure intuitively, we plot the two-dimensional ($z_Q, r_Q$) diagrams of these two types of jets in Fig.~\ref{fig:pp-zrQ} and Fig.~\ref{fig:pp-zrg} in p+p collisions at $\sqrt{s}=$ 5.02 TeV for both c-jet and b-jet. Here $z_Q=p_T^Q/p_T^{\rm jet}$ denotes the transverse momentum fraction carried by heavy quarks in jets, and $r_Q=\sqrt{(\phi_Q-\phi_{\rm jet})^2+(\eta_Q-\eta_{\rm jet})^2}$ the radial distance of heavy quarks to the jet axis. These two observables, $z_Q$ and $r_Q$, quantify the energy dominance of heavy quarks in jets and their angular location in the jet cone. For $Q\rightarrow Q$-jet of both c-jet and b-jet, one can see that most heavy quarks carry above $80\%$ of the jet momentum; simultaneously, their moving directions collimate with the jet axis. Understandably, heavy quarks will still dominate the jet's momentum even after soft shower evolution. However, the situation looks quite different for $g\rightarrow Q$-jet in the lower panels, where heavy quarks are more dispersive to distribute in the ($z_Q, r_Q$) diagrams. We observe a banded region, especially for $g\rightarrow c$-jet, which is distinctly different from that of $c\rightarrow c$-jet. Plenty of heavy quarks produced by the gluon splitting carry smaller energy fractions and are located in larger radii in jets. In the remainder of this paper, we will show that the large fraction of $g\rightarrow Q$-jet and its distinct substructure compared to $Q\rightarrow Q$-jet will play critical roles in the energy loss and yield suppression of HQ jets in high-energy nucleus-nucleus collisions.

%%%%%%%%%%%%%%%%%%%%%%%%%%%%%%%%%%%%%%%%%%%%%%%%%%%%%%%%%%%%%%%%%%%%%
%%%%%%%%%%%%%%%%%%%%%%%%%%%%%%%%%%%%%%%%%%%%%%%%%%%%%%%%%%%%%%%%%%%%%
%%%%%%%%%%%%%%%%%%%%%%%%%%%%%%%%%%%%%%%%%%%%%%%%%%%%%%%%%%%%%%%%%%%%%
%%%%%%%%%%%%%%%%%%%%%%%%%%%%%%%%%%%%%%%%%%%%%%%%%%%%%%%%%%%%%%%%%%%%%

\section{Jet transport in the quark-gluon plasma}
\label{sec:framework}

Due to the large mass ($M_Q\gg T$), the heavy quarks are effective hard probes to the properties of the hot and dense QCD matter formed in high-energy nuclear collisions. In this study, we utilize the p+p events produced by SHERPA as input of the transport model driven by the modified Langevin equations~\cite{Moore:2004tg, Rapp:2009my, He:2013zua, Katz:2019fkc, Li:2020umn, Cao:2013ita} to estimate the nuclear modification effect of the HQ jet production in A+A collisions.

\begin{eqnarray}
{\rm d}x_j&=&\frac{p_j}{E}{\rm d} t \\
{\rm d}p_j&=&-\Gamma p_j{\rm d} t+\sqrt{{\rm d} t}C_{jk}(|\mathbf{p}+\xi{\rm d}\mathbf{p}|)\rho_k-p^{\rm g}_j
\label{eq:lang2}
\end{eqnarray}

Eq.~(1) describes the spatial position update of the traversing heavy quarks in the medium, where ${\rm d} t=0.1$ fm is the time step of our simulation. Meanwhile, Eq.~(2) simulates their in-medium energy loss with index $j, k=1,2,3$, where the three terms on the right-hand side denote the drag term, the thermal stochastic term, and the radiative correction term, respectively. The drag term represents the collisional energy loss of heavy quarks, where the drag coefficient $\Gamma$ controls the strength of the energy loss. Besides, the thermal stochastic term means the mass of random kicks suffered on heavy quarks caused by the thermal quasi-particles in the hot and dense QCD matter, where the  white $\rho_k$ obeys a standard normal distribution, $P({\rho})=(2\pi)^{-3/2}e^{\rho^2/2}$. The momentum argument of the covariance matrix $C_{jk}$ is a function of the momentum diffusion coefficients at the longitudinal ($\kappa_{||}$) and the transverse ($\kappa_{\perp}$) direction~\cite{Beraudo:2009pe}.

\begin{eqnarray}
C_{jk}=\sqrt{\kappa_{||}}\frac{p_jp_k}{\vec{p}^{\, 2}}+\sqrt{\kappa_{\perp}}(\delta_{jk}-\frac{p_jp_k}{\vec{p}^{\, 2}})
\label{eq:c12}
\end{eqnarray}

 Note here we have chosen $\xi=0$ for the pre-point (Ito) realization of the stochastic integral of $C_{jk}(|\mathbf{p}+\xi{\rm d}\mathbf{p}|)$~\cite{Ito:1951aaa}. Additionally, by assuming $\kappa$ is isotropic for heavy quarks $\kappa_{\perp}=\kappa_{||}=\kappa$, one can obtain a simple expression $C_{jk}=\sqrt{\kappa}\delta_{jk}$. Then the momentum diffusion coefficient $\kappa$ could be related to the drag coefficient $\Gamma$ by the relativistic Einstein relation $\Gamma=\kappa/2ET$. Note that the momentum diffusion coefficient $\kappa$ of heavy quarks can also be converted into another dimensionless form $2\pi TD_s$ in coordinate space with the relation $D_s=2T^2/\kappa$. Note that, to accurately determine the temperature and momentum dependence of the diffusion coefficient of heavy quarks ($\kappa$ or $D_s$), some elaborate and significant efforts from the heavy-ion community have been made in recent years~\cite{Xu:2018gux, Cao:2017hhk, Francis:2015daa, Ding:2012sp, Altenkort:2023eav}. More detailed reviews can be find in Refs.~\cite{Rapp:2018qla, Apolinario:2022vzg}. At each time step, we boost heavy quarks to the local rest frame of the expanding medium to update the four-momentum by the Lorentz transformation and then boost them back to the laboratory frame to update the spatial position. The last correction term $-p^g_j$ corresponds to the momentum recoil of the medium-induced radiated gluon, based on the gluon spectrum calculated by the higher-twist approach~\cite{Guo:2000nz, Zhang:2003yn,Zhang:2003wk, Majumder:2009ge},

\begin{eqnarray}
\frac{dN}{ dxdk^{2}_{\perp}dt}=\frac{2\alpha_{s}C_sP(x)\hat{q}}{\pi k^{4}_{\perp}}\sin^2(\frac{t-t_i}{2\tau_f})(\frac{k^2_{\perp}}{k^2_{\perp}+x^2M^2})^4,
\label{eq:dndxk}
\end{eqnarray}
where $x$ and $k_\perp$ are the energy fraction and the transverse momentum carried by the radiated gluon. $C_s$ is the quadratic Casimir in the color representation, and $P(x)$ the splitting function, $\tau_f=2Ex(1-x)/(k^2_\perp+x^2M^2)$ denotes the gluon formation time to take into account the Landau-Pomeranchuk-Migdal (LPM) effects \cite{Wang:1994fx,Zakharov:1996fv}. $\hat{q}=q_0(T/T_0)^3p_{\mu}u^{\mu}/E$ is the jet transport parameter~\cite{Chen:2010te}, where $T_0$ is the highest temperature in the most central A+A collisions and $u^{\mu}$ the velocity of the medium cell where the heavy quark locates. To take into account the fluctuation of medium-induced gluon radiation, we assume that the number of the radiated gluon during a time step $\Delta t$ obeys Possion distribution $f(n)=\lambda^{n}e^{-\lambda}/{n!}$, where the parameter $\lambda$ denotes the mean number of the radiated gluon which can be calculated by integrating Eq.~(\ref{eq:dndxk}). Once the radiation number $n$ is sampled, the radiated gluon's four-momentum can be further sampled by Eq.~(\ref{eq:dndxk}) one by one. As discussed above, the momentum update of heavy quarks is driven by the Eq.~(\ref{eq:lang2}). Besides, for calculating the full-jet observable, it is essential to consider the energy loss of the light partons (light quark and gluon) inside the jet cone. As an effective treatment, the elastic energy loss of light partons is estimated by the pQCD calculation at HTL approximation~\cite{Neufeld:2010xi} while the inelastic part is by the higher-twist formalisms.

The initial spacial production vertex of jets in nucleus-nucleus collisions is sampled based on the MC-Glauber model~\cite{Miller:2007ri}. During jet propagation, the velocity and temperature of the expanding QGP medium are provided by the CLVisc hydrodynamic model~\cite{Pang:2016igs}. In general, it is assumed that the partonic energy loss ceases at the hadronic phase, namely the local temperature below $T_c=0.165$~GeV. In this work, we neglect the initial fluctuation of the bulk medium due to its small influence on the average energy loss of high-$p_T$ jet~\cite{Betz:2011tu, Renk:2011qi, Cao:2014fna}. However, it should be noted that such initial fluctuation of hydrodynamics may be critical for the simultaneous description of the $R_{AA}$ and $v_2$ of heavy-flavor hadrons at low $p_T$ region~\cite{Noronha-Hostler:2016eow, Prado:2016szr}. Note that, based on the isotropic approximation of the QGP medium, $\kappa$ and $\hat{q}$ of a high-energy heavy quark can be related with a concise expression $\kappa=\hat{q}/2$ \cite{Cao:2013ita, Rapp:2018qla, Cao:2018ews}. In recent years, the SHELL transport model has been successfully employed to study the medium modifications of both light- and heavy-flavor jet in high-energy nuclear collisions \cite{Dai:2018mhw, Wang:2019xey, Wang:2020ukj, Wang:2021jgm, Li:2022tcr, Wang:2023eer, Wang:2023udp, Li:2024uzk, Wang:2024plm, Li:2024pfi}.

\begin{widetext}

\begin{figure}[!t]
\begin{center}
\vspace*{0.1in}
\hspace*{-0.3in}
\includegraphics[width=6in,angle=0]{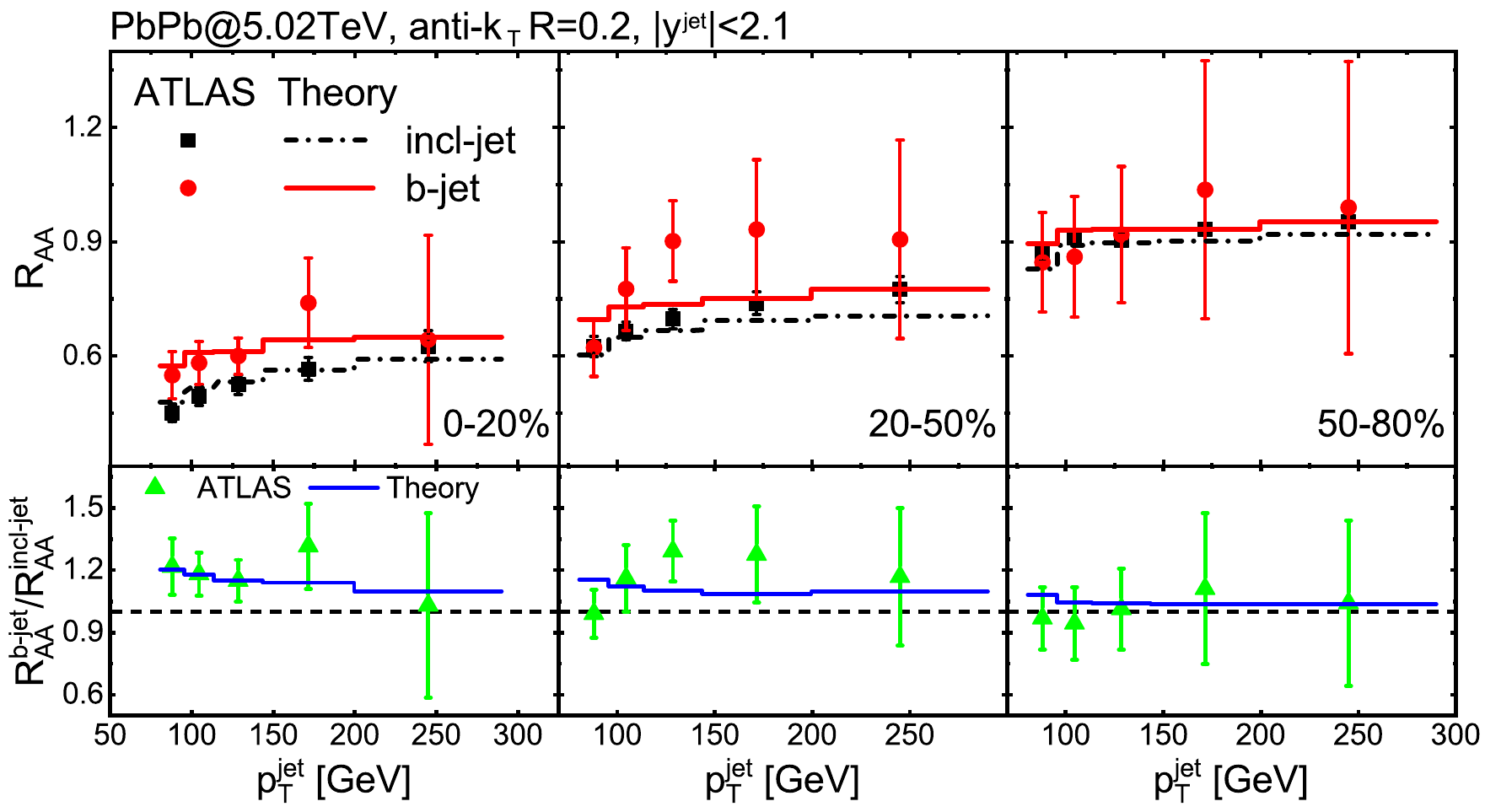}
\vspace*{0in}
\caption{(Color online) The nuclear modification factor $R_{AA}$ of b-jet and inclusive jet in $0-20\%$, $20-50\%$ and $50-80\%$ Pb+Pb collisions at $\sqrt{s_{NN}}=$ 5.02 TeV compared with ATLAS data (upper panels), as well as their ratio $R_{\rm AA}^{\rm b\text{-}jet}/R_{\rm AA}^{\rm incl\text{-}jet}$ (lower panels).}
\vspace*{-0.2in}
\label{fig:raa-ib}
\end{center}
\end{figure}

\end{widetext}

%%%%%%%%%%%%%%%%%%%%%%%%%%%%%%%%%%%%%%%%%%%%%%%%%%%%%%%%%%%%%%%%%%%%%
%%%%%%%%%%%%%%%%%%%%%%%%%%%%%%%%%%%%%%%%%%%%%%%%%%%%%%%%%%%%%%%%%%%%%
%%%%%%%%%%%%%%%%%%%%%%%%%%%%%%%%%%%%%%%%%%%%%%%%%%%%%%%%%%%%%%%%%%%%%
%%%%%%%%%%%%%%%%%%%%%%%%%%%%%%%%%%%%%%%%%%%%%%%%%%%%%%%%%%%%%%%%%%%%%
\section{Results and discussions}
\label{sec:res}
\begin{figure}[!t]
\begin{center}
\includegraphics[width=3.in,angle=0]{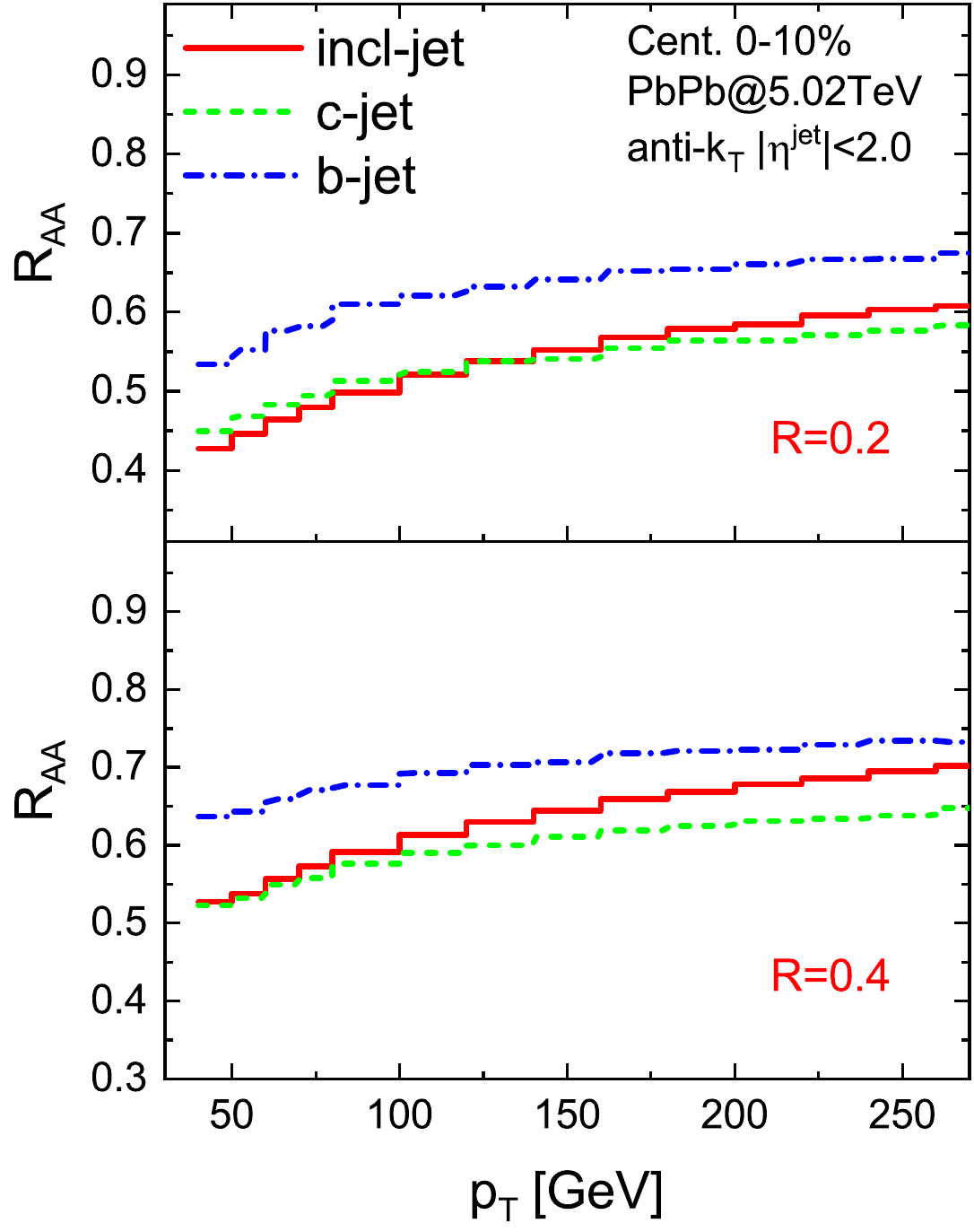}
\vspace*{0in}
\caption{(Color online) The nuclear modification factor $R_{AA}$ of both c-jet, b-jet and inclusive jet in central $0-10\%$ Pb+Pb collisions at $\sqrt{s_{NN}}=$ 5.02 TeV for different jet-cone size (R = 0.2, 0.4).}
\label{fig:raa-icb}
\end{center}
\end{figure}

Note that $\hat{q}$ is the only parameter in our framework that controls the interaction strength for both light and heavy-flavor partons. Firstly, we perform a $\chi^2$ fit of the recent inclusive-jet and b-jet $R_{AA}$ data measured by the ATLAS collaboration \cite{ATLAS:2022agz} to fix the model parameter $\hat{q}$, where the jet $R_{AA}$ is conventionally defined as,

\begin{eqnarray}
R_{AA}=\frac{1}{\left\langle N_{\rm bin}^{\rm AA} \right\rangle}\frac{d\sigma^{\rm AA}/\rm dydp_T}{d\sigma^{\rm pp}/\rm dydp_T}
\label{eq:raa}
\end{eqnarray}
the scaling factor $\left\langle N_{\rm bin}^{\rm AA} \right\rangle$ denotes the number of binary nucleon-nucleon collisions in A+A estimated with the Glauber model~\cite{Miller:2007ri}. The $\chi^2$ fit of the ATLAS data \cite{ATLAS:2022agz} gives the optimal value of $q_0=$ 0.9 GeV$^2$/fm with $\chi^2=$1.29. With this configuration, we show our model calculations of b-jet and inclusive jet $R_{AA}$ compared with the data in Fig.~\ref{fig:raa-ib}, which shows good agreement for both the $R_{AA}$ magnitudes and their ratio $R_{AA}^{\textit{\rm b-jet}}/R_{AA}^{\textit{\rm incl-jet}}$ from central to peripheral Pb+Pb collisions at $\sqrt{s_{NN}}=$ 5.02TeV. The b-jets suffer a more moderate yield suppression than the inclusive jet, which may indicate that the b-jet loses less energy in QGP than the light-flavor jet. However, as we have discussed in the \ref{sec:ppbaseline}, the HQ-initiated jet is not the dominant contribution for c-jet and b-jet samples as discussed in Sec.~\ref{sec:ppbaseline}. The mixture of components in HQ jets complicates understanding the mass effect of jet energy loss. As shown in Fig.~\ref{fig:raa-icb}, we estimate the nuclear modification factor $R_{AA}$ of both c-jet, b-jet and inclusive jet in central $0-10\%$ Pb+Pb collisions at $\sqrt{s_{NN}}=$ 5.02 TeV for different jet-cone size (R = 0.2, 0.4), at $30<p_T<260$ GeV. These calculations are performed at the parton level, and HQ jets (c-jet and b-jet) are defined as jets containing at least one heavy quark inside the jet cone with $p_T^Q>$ 3 GeV. Except that the b-jet has the largest $R_{AA}$, it is surprising that the c-jet has comparable $R_{AA}$ with inclusive jet for R = 0.2. In addition, for larger jet cone R = 0.4, we find that the yield suppression of c-jet is stronger than that of inclusive jet at $p_T>$ 150 GeV. We notice that similar results are also obtained by the LIDO model \cite{Ke:2020nsm}.

\begin{figure}[!t]
\begin{center}
\includegraphics[width=2.97in,angle=0]{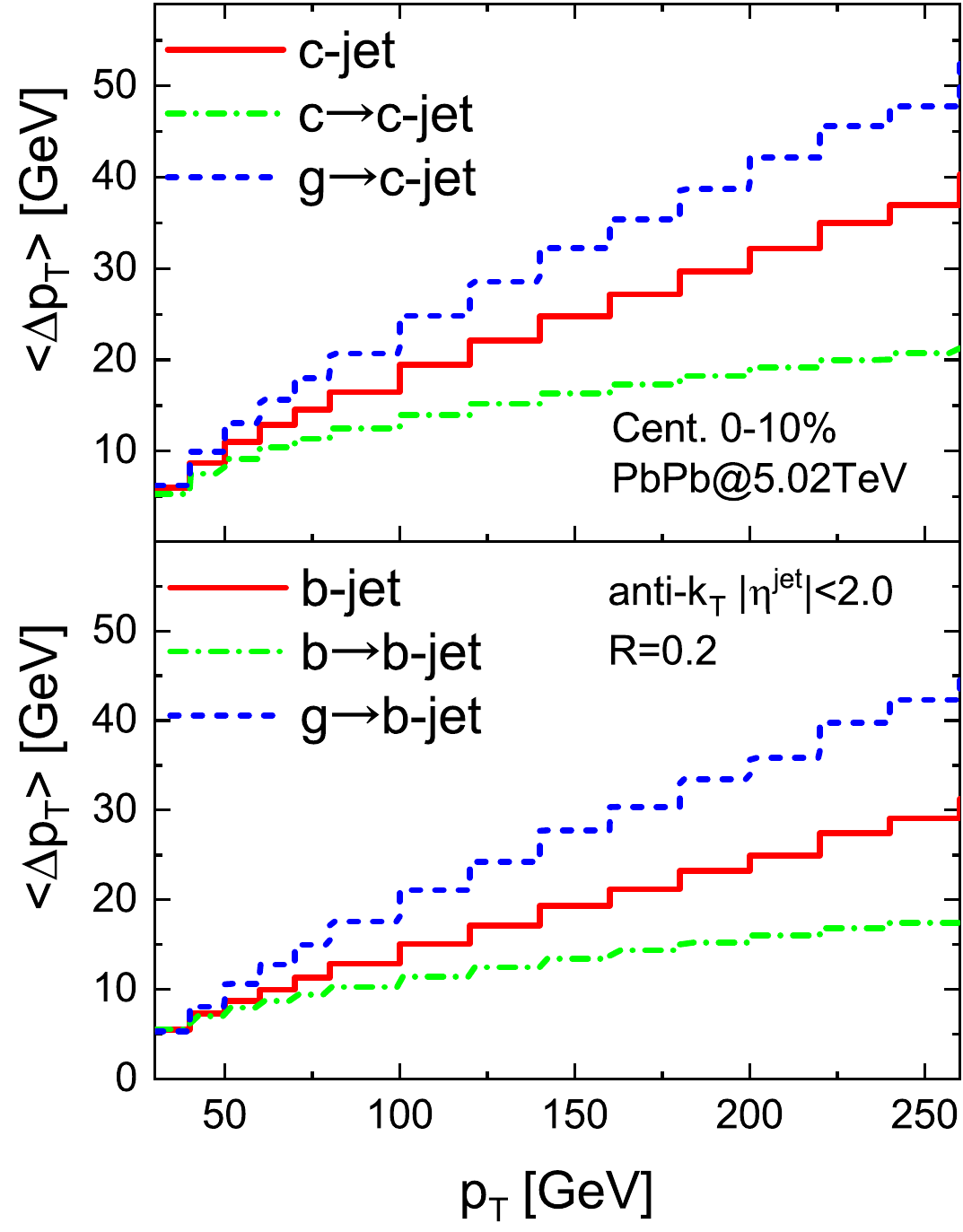}
\vspace*{0in}
\caption{(Color online) The averaged transverse jet energy loss ($\left\langle\Delta p_T\right\rangle$) of $Q \rightarrow Q$-jet and $g \rightarrow Q$-jet in $0-10\%$ Pb+Pb collisions at $\sqrt{s_{NN}}=$ 5.02 TeV, both for c-jet and b-jet.}
\label{fig:dpt-cb-sub}
\end{center}
\end{figure}

In Fig.~\ref{fig:dpt-cb-sub} we estimate the averaged jet energy loss ($\Delta p_T$) of $Q \rightarrow Q$-jet and $g \rightarrow Q$-jet by tracking their propagation in $0-10\%$ Pb+Pb collisions at $\sqrt{s_{NN}}=$ 5.02 TeV for c-jet (upper) and b-jet (lower). First, the total c-jet $\Delta p_T$ seems more significant than that of b-jet with the same initial $p_T$. Second, by comparing the $\Delta p_T$ of $c \rightarrow c$-jet and $b \rightarrow b$-jet, it is easy to see that Q-jets initiated by the bottom lose less energy than that by charm, which is the direct embodiment of the mass hierarchy of quark energy loss. Finally, we find that $g \rightarrow Q$-jet loses much more energy than $Q \rightarrow Q$-jet for both c-jet and b-jet. This work employs the p+p events after a full vacuum parton shower as input to simulate the in-medium energy loss. This treatment assumes all heavy quarks were created before the QGP was formed. Hence, the different energy losses of $Q \rightarrow Q$-jet and $g \rightarrow Q$-jet may only result from their initial features. As we have discussed in Fig.~\ref{fig:pp-zrQ} and Fig.~\ref{fig:pp-zrg} at the end of Sec.~\ref{sec:ppbaseline}, the $g \rightarrow Q$-jet generally has more dispersive ($z_Q, r_Q$) distribution compared to $Q \rightarrow Q$-jet, namely plenty of heavy quarks locate in smaller $z_Q$ and larger $r_Q$ region. Larger $r_Q$ makes the lost energy from heavy quarks easier to dissipate outside the jet cone. It is consistent with the results of the recent measurements of the ATLAS, ALICE, and CMS collaborations~\cite{ATLAS:2022vii, ALargeIonColliderExperiment:2021mqf, ALICE:2018dxf, ALICE:2023dwg, ATLAS:2023hso, CMS:2024zjn}, which indicate that the more dispersive the jet structure the more energy it will lose in the QGP. Moreover, jets produced by $g \rightarrow Q\bar{Q}$ processes usually contain two heavy quarks inside the jet cone, whose energy dissipation in the QGP is more efficient compared to the $Q \rightarrow Q$-jet.

\begin{figure}[!t]
\begin{center}
%\hspace*{-0.1in}
\includegraphics[width=3.0in,angle=0]{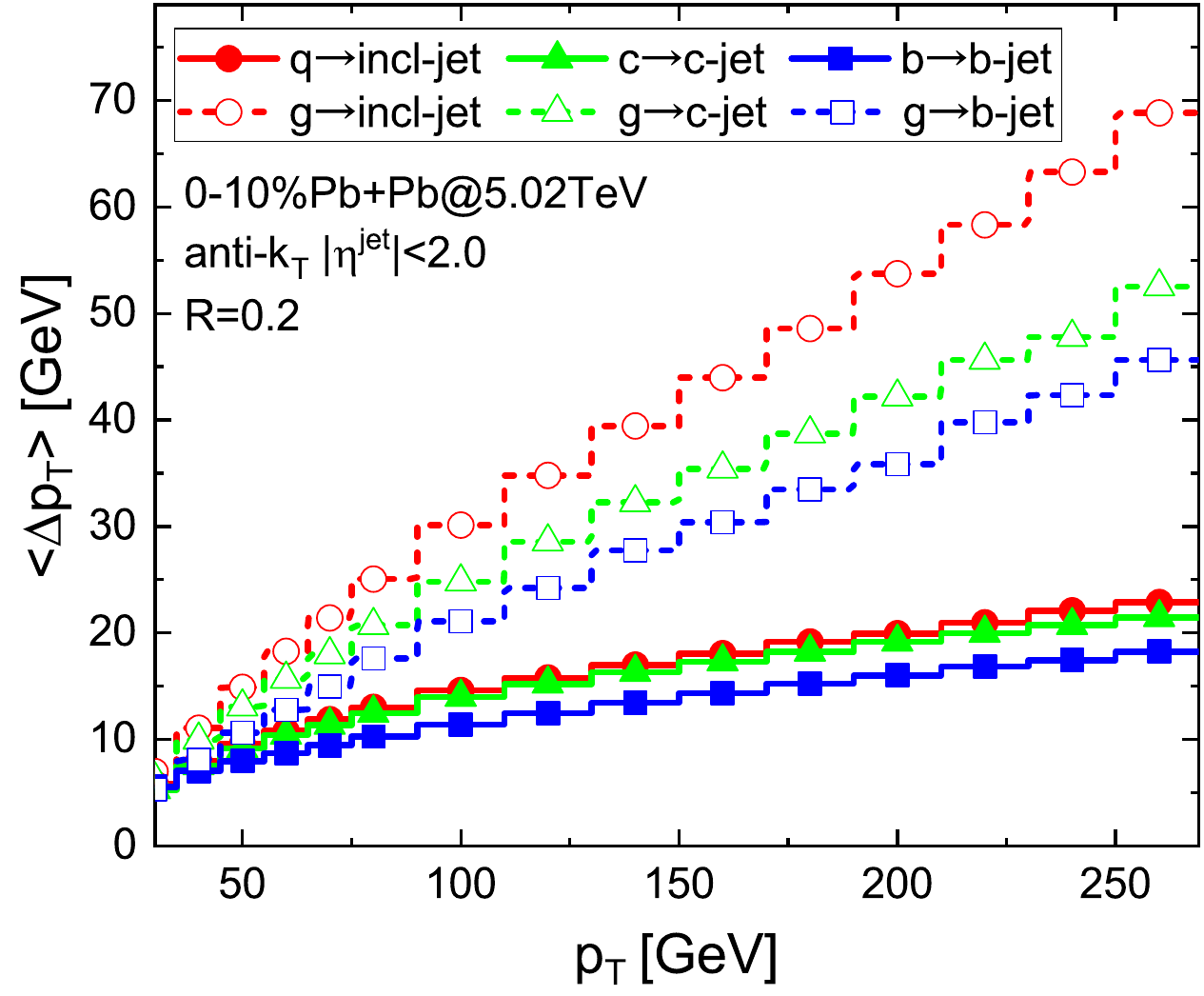}
\vspace*{0in}
\caption{(Color online) The averaged transverse energy loss ($\left\langle\Delta p_T\right\rangle$) for the six kinds of jets, $q \rightarrow \rm incl$-jet, $g \rightarrow \rm incl$-jet, $c \rightarrow c$-jet, $g \rightarrow c$-jet, $b \rightarrow b$-jet and $g \rightarrow b$-jet, in $0-10\%$ Pb+Pb collisions at $\sqrt{s_{NN}}=$ 5.02 TeV.}
\label{fig:dpt-icbR2}
\end{center}
\end{figure}

\begin{figure}[!t]
\begin{center}
%\hspace{-0.1in}
\includegraphics[width=3.2in,angle=0]{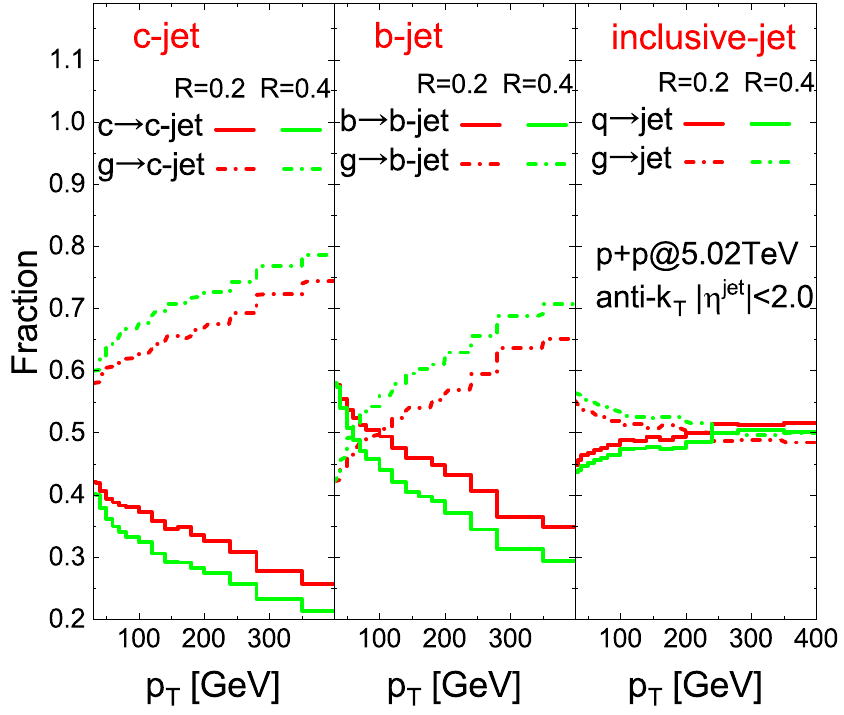}
\vspace*{-0.1in}
\caption{(Color online) The fractional contributions of $c \rightarrow c$-jet and $g \rightarrow c$-jet to the total c-jet, $b \rightarrow b$-jet and $g \rightarrow b$-jet to the total b-jet, $q \rightarrow \rm incl$-jet and $g \rightarrow \rm incl$-jet to the inclusive jet differential cross section versus $p_T$ in p+p collisions at $\sqrt{s}=$ 5.02 TeV, as jet-cone size varies from 0.2 to 0.4.}
\label{fig:frac-icbR}
\end{center}
\end{figure}

On the other hand, we can have an overall comparison of the $\left\langle\Delta p_T\right\rangle$ for the six kinds of jets ($q \rightarrow \rm incl$-jet, $g \rightarrow \rm incl$-jet, $c \rightarrow c$-jet, $g \rightarrow c$-jet, $b \rightarrow b$-jet and $g \rightarrow b$-jet) in $0-10\%$ Pb+Pb collisions at $\sqrt{s_{NN}}=$ 5.02 TeV as shown in Fig.~\ref{fig:dpt-icbR2}, to test the flavor dependence of jet energy loss intuitively and systematically. Firstly, comparing the energy loss of the jet initiated by the parton with different flavors is essential. Excluding the contribution from gluon splitting processes, we find that the jet energy loss obeys the order $\Delta E_{g\rightarrow \textit{\rm incl-jet}}>\Delta E_{q\rightarrow \textit{\rm incl-jet}}\gtrsim\Delta E_{c\rightarrow \textit{c-\rm jet}}>\Delta E_{b\rightarrow \textit{b-\rm jet}}$, in line with the flavor-dependent parton energy loss expectation. We notice that $b \rightarrow b$-jet shows a noticeable mass effect of energy loss compared to $q \rightarrow $incl-jet, while the $c \rightarrow c$-jet behaved more like a light quark jet at $p_T>$ 50 GeV. Besides, it is also interesting to compare the energy loss of $g \rightarrow $incl-jet with that of $g \rightarrow Q$-jet. They are all initiated by the high-energy gluon, but the latter traverses the QGP as a $Q\bar{Q}$ pair inside the jet. The energy loss of $g \rightarrow c$-jet is smaller than that of $g \rightarrow \rm jet$ but visibly larger than $q \rightarrow \rm incl$-jet. It indicates that the HQ jets from the gluon splitting lose more energy in the QGP than the light-quark jets. Therefore, the significant contribution of $g \rightarrow Q$-jet may be the critical point to understand the similar $R_{AA}$ of c-jet and inclusive jet shown in Fig. \ref{fig:raa-icb}. In other words, due to the large fraction of $g \rightarrow c$-jet components, the averaged energy loss of c-jet can be comparable to that of the inclusive jet. In Fig.~\ref{fig:frac-icbR}, we estimate the fractions of $Q \rightarrow Q$-jet and $g \rightarrow Q$-jet for c-jet (left panel) and b-jet (middle panel) versus jet $p_T$ (left panel) with R = 0.2 and R = 0.4. It is observed that the fractions of $g \rightarrow Q$-jet elevate when R varies from 0.2 to 0.4, and accordingly, that of $Q \rightarrow Q$-jet decreases for both c-jet and b-jet. For larger jet cones, the enhanced fraction of $Q \rightarrow Q$-jet increases the average energy loss of HQ jets. Additionally, the fractions of quark- and gluon-jet in the inclusive jet sample are shown in the right panel, and they turn out to be insensitive to the jet-cone size. It could explain that the c-jet $R_{AA}$ can be smaller than that of the inclusive jet for R = 0.4, as shown in Fig. \ref{fig:raa-icb}.

 So far, we have discussed the jet energy loss of c-jet, b-jet, and inclusive jet by systematically analyzing their component features and fraction variations. We point out that the energy loss of $g \rightarrow Q$-jet is significantly more pronounced than that of $Q \rightarrow Q$-jet in nucleus-nucleus collisions. Our calculations indicate that $g \rightarrow Q$-jet behaves like a gluon-jet but not a heavy quark jet. More efforts in future experimental measurements on $g \rightarrow Q$-jet and $Q \rightarrow Q$-jet can be helpful and essential to address the flavor/mass dependence of jet energy loss. However, the presuppose of these measurements is that the reconstructed Q-jets from different channels ($Q \rightarrow Q$-jet and $g \rightarrow Q$-jet) can be identified effectively in the experiment. For this reason, we will show the selection methods to separate the two processes $g \rightarrow Q$-jet and $Q \rightarrow Q$-jet and also estimate the purity of the jet sample selected by these strategies.

 As we mentioned in Sec.~\ref{sec:ppbaseline}, the $Q\bar{Q}$ pairs produced by $g \rightarrow Q$-jet usually have a narrow opening angle while the one from hard scattering is usually ``back-to-back'' in the azimuthal plane. We design the following strategies to select the high-purity sample of $g \rightarrow Q$-jet and $Q \rightarrow Q$-jet, respectively.

 \begin{itemize}
\item {\bf Strategy-1} for $g \rightarrow Q$-jet: Selecting jets containing two heavy quarks inside the jet-cone. The heavy quark should have $p_T^Q>$2 GeV.
\item {\bf Strategy-2} for $Q \rightarrow Q$-jet: Selecting jets containing only one heavy quark inside the jet-cone. Moreover, the selected candidates should have a recoiled HQ jet partner with $p_T>$10 GeV, and their angle separation in azimuth plane satisfy $\Delta \phi_{12}>2/3\pi$.
\end{itemize}

\begin{figure}[!t]
\begin{center}
\includegraphics[width=3.in,angle=0]{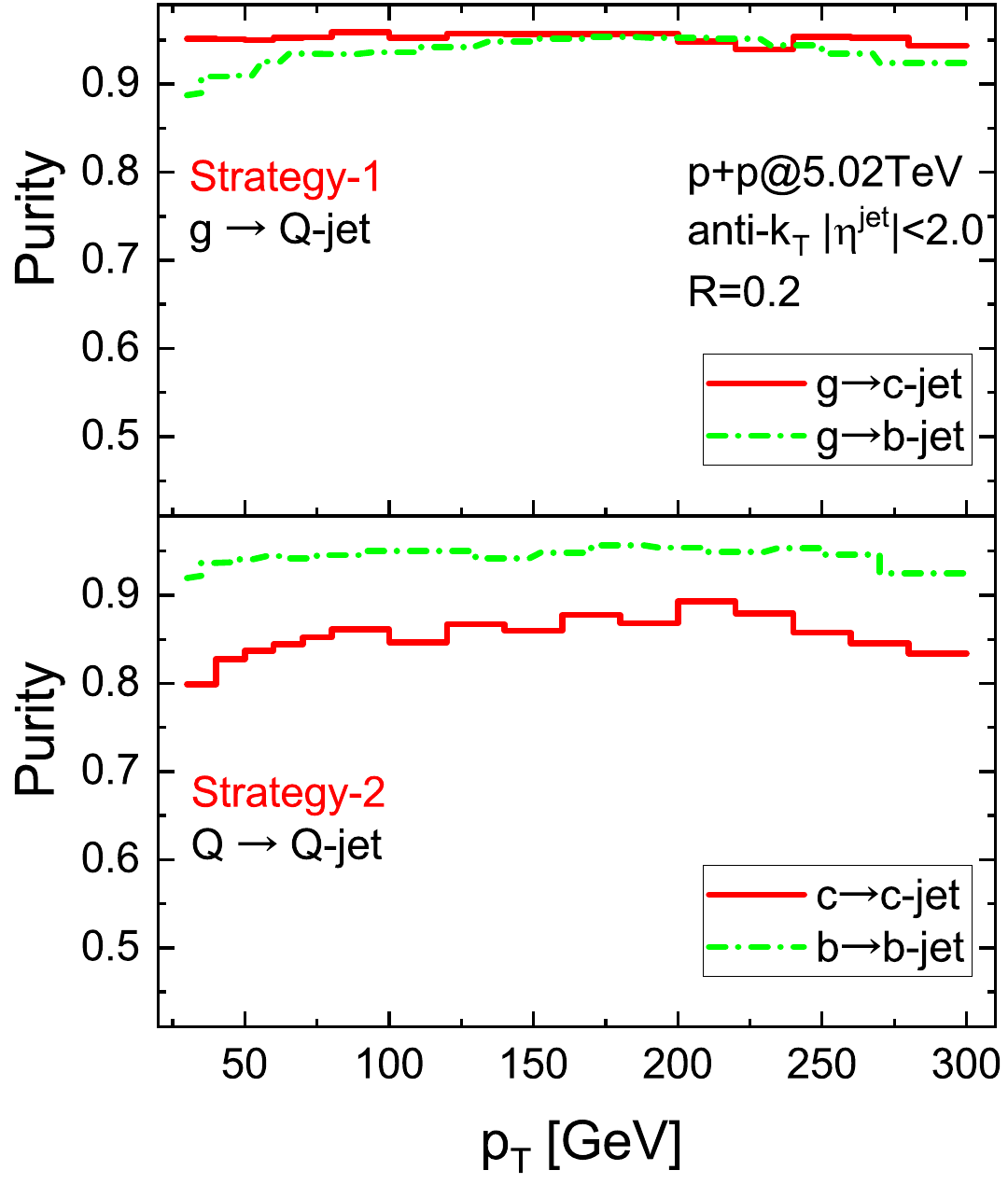}
\vspace*{0in}
\caption{(Color online) Purities of the selected of $g \rightarrow Q$-jet and $Q \rightarrow Q$-jet samples for c-jet and b-jet in p+p collisions at $\sqrt{s_{NN}}=$ 5.02 TeV.}
\label{fig:frac1Q}
\end{center}
\end{figure}

To test the performance of the selection strategies, in Fig.~\ref{fig:frac1Q} we show the purities of the selected $g \rightarrow Q$-jet and $Q \rightarrow Q$-jet samples in p+p collisions at $\sqrt{s_{NN}}=$ 5.02 TeV for both c-jet and b-jet, where the purity is defined as the fraction of the target process in the total selected jet sample,

 \begin{eqnarray}
{\rm Purity}=\frac{d\sigma/dp_T[{\rm target \: process}]}{d\sigma/dp_T[{\rm selected \: jet \: sample}]}.
\label{eq:SE}
\end{eqnarray}
In the upper panel of Fig.~\ref{fig:frac1Q}, the purity of the selected $g \rightarrow Q$-jet sample is above 0.9 for both c-jet and b-jet. In the lower panel, the purity of the selected $Q \rightarrow Q$-jet sample is above 0.8 for c-jet and 0.9 for b-jet. Moreover, the selection strategies have also been tested in A+A collisions and shown similar satisfactory performance.

\begin{figure}[!t]
\begin{center}
\includegraphics[width=3.2in,angle=0]{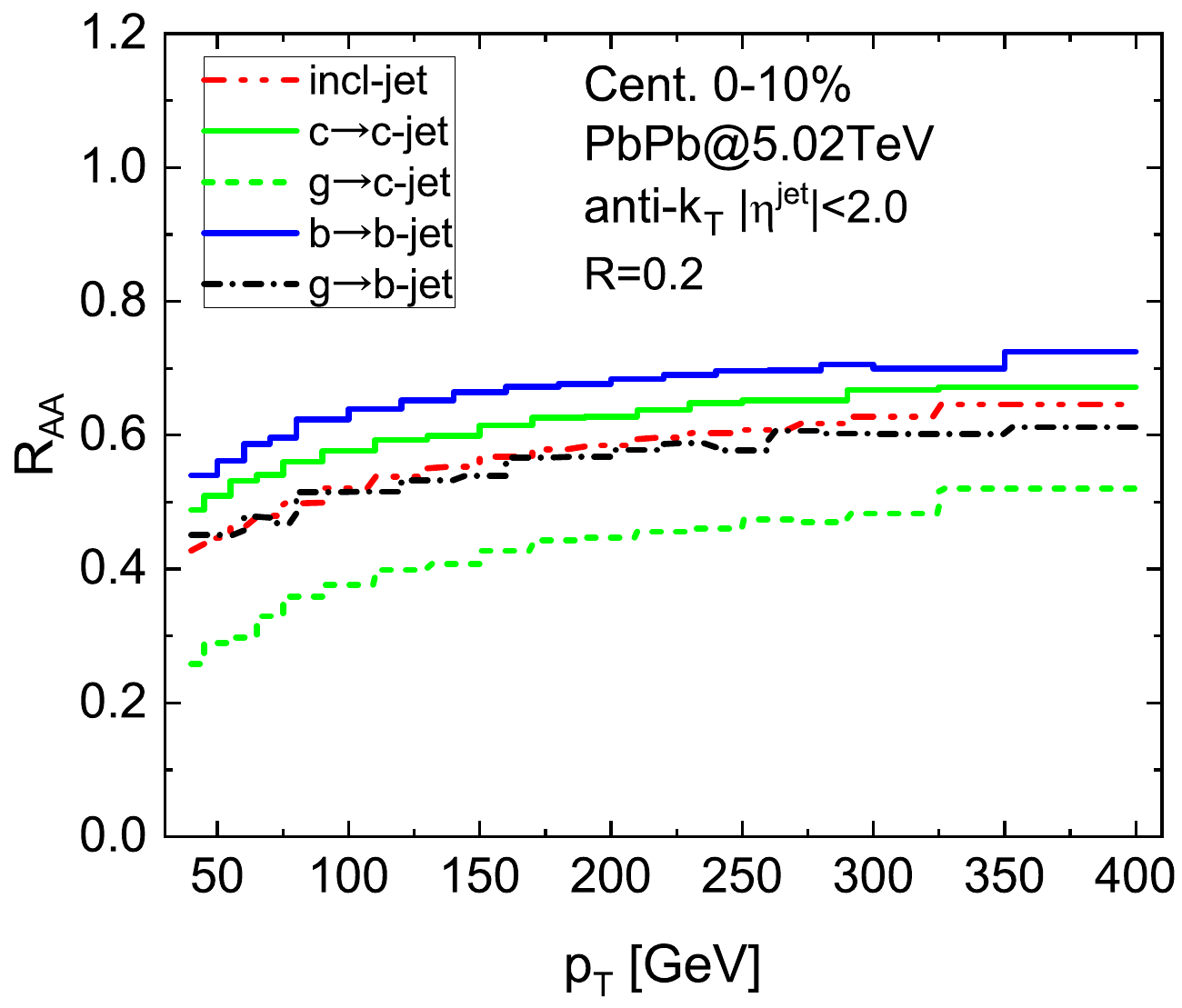}
\caption{(Color online) $R_{AA}$ of the selected $c \rightarrow c$-jet, $g \rightarrow c$-jet, $b \rightarrow b$-jet and $g \rightarrow b$-jet compared to that of the inclusive jet in $0-10\%$ Pb+Pb collisions at $\sqrt{s_{NN}}=$ 5.02 TeV.}
\label{fig:RAA-12Q}
\end{center}
\end{figure}

After establishing the effective strategies to select the $g \rightarrow Q$-jet and $Q \rightarrow Q$-jet, we can directly compare their yield suppression in A+A collisions with the inclusive jet. Although it is still difficult to isolate the gluon jet and quark jet in the experiment, the comparison between $g \rightarrow Q$-jet, $Q \rightarrow Q$-jet, and inclusive jet may provide a unique chance to capture the flavor/mass dependence of jet quenching, which can be tested in experiment accessibly. In Fig.~\ref{fig:RAA-12Q}, we present the calculations of $R_{AA}$ of selected $c \rightarrow c$-jet, $g \rightarrow c$-jet, $b \rightarrow b$-jet and $g \rightarrow b$-jet compared to that of inclusive jet in $0-10\%$ Pb+Pb collisions at $\sqrt{s_{NN}}=$ 5.02 TeV. Note here that $c \rightarrow c$-jet, $g \rightarrow c$-jet, $b \rightarrow b$-jet and $g \rightarrow b$-jet denote the selected samples with the above mentioned strategies. It is clear to see that jets initiated by heavy quarks obey the order $R_{AA}^{\textit{\rm b-jet}}>R_{AA}^{\textit{\rm c-jet}}>R_{AA}^{\textit{\rm incl-jet}}$. Namely, by effectively isolating the $Q \rightarrow Q$-jet processes, we predict that the mass hierarchy of energy loss at jet level ($\Delta E_{\textit{\rm incl-jet}}>\Delta E_{\textit{\rm c-jet}}>\Delta E_{\textit{\rm b-jet}}$) is held in nucleus-nucleus collisions. On the other hand, we observe that the yield suppression of $g \rightarrow c$-jet is much stronger than that of inclusive jet. As for $g \rightarrow b$-jet, its yield suppression is obviously stronger than that of $b \rightarrow b$-jet and $c \rightarrow c$-jet. It is also found that at high $p_T$, $R_{AA}$ of $g \rightarrow b$-jet is slightly lower than that of inclusive jet while their values are close at lower $p_T$. That may result from the dispersive structure of the b-jets produced by the $g \rightarrow b$-jet process which can lead to more energy loss in the QGP compared to $b \rightarrow b$-jet, similar to the case of c-jet.

It should be noted that different from the treatment in Ref. \cite{Xing:2019xae}, which considers the fragmentation function of high-energy gluon into the heavy-flavor hadron in the hadronization process, in this work, we simulate the vacuum splitting of $g \rightarrow Q$-jet before the formation of the QGP medium. The results obtained with different treatments are consistent, indicating that the heavy flavors from gluon splitting suffer a stronger quenching effect than those from the hard scattering in heavy-ion collisions. Our next plan is to investigate the influence of considering the concrete $g \rightarrow Q$-jet splitting time in the QGP medium to the energy loss and substructure modifications of HQ jets, while some recent exploratory studies in this direction should be helpful \cite{JETSCAPE:2022hcb, Zhang:2022ctd, Modarresi-Yazdi:2024vfh}. Moreover, we notice that the recent studies have extracted the gluon energy loss by $J/\Psi$ production in Pb+Pb collisions \cite{Zhang:2022rby} because over $80\%$ $J/\Psi$ are mainly produced by the gluon fragmentation at high $p_T$. From this point of view, measurement on the $g \rightarrow c$-jet production in A+A collisions could similarly build a bridge for understanding the energy loss of the gluon jet.

%%%%%%%%%%%%%%%%%%%%%%%%%%%%%%%%%%%%%%%%%%%%%%%%%%%%%%%%%%%%%%%%%%%%%
%%%%%%%%%%%%%%%%%%%%%%%%%%%%%%%%%%%%%%%%%%%%%%%%%%%%%%%%%%%%%%%%%%%%%
%%%%%%%%%%%%%%%%%%%%%%%%%%%%%%%%%%%%%%%%%%%%%%%%%%%%%%%%%%%%%%%%%%%%%
%%%%%%%%%%%%%%%%%%%%%%%%%%%%%%%%%%%%%%%%%%%%%%%%%%%%%%%%%%%%%%%%%%%%%
\section{Summary}
\label{sec:sum}

This paper systematically studies the yield suppression of HQ jets in nucleus-nucleus collisions relative to the p+p. It focuses on the energy loss of HQ jets produced by different production mechanisms. We use the Monte Carlo event generator SHERPA to generate the p+p baseline, which matches the NLO hard QCD processes with the resummation of the parton shower. The in-medium evolution of the HQ jets is described by the SHELL transport model, which considers the elastic and inelastic energy loss. One can see that the $g \rightarrow Q$-jet process significantly contributes to the HQ jet production at high $p_T$ and shows more dispersive structures compared to the $Q \rightarrow Q$-jet in p+p collisions. In Pb+Pb collisions at $\sqrt{s_{NN}}=$ 5.02 TeV, our calculations give decent descriptions of the inclusive jet and b-jet $R_{AA}$ measured by the ATLAS collaboration, which suggests a remarkable hint of the mass effect of jet energy loss. Due to the dispersive substructure, we find that $g \rightarrow Q$-jet loses more energy than the $Q \rightarrow Q$-jet in the same collision system. Furthermore, due to the dominant contribution of $g \rightarrow c$-jet, the $R_{AA}$ of c-jet will be comparable or even smaller than that of inclusive jet. We propose the event selection strategies based on their topological features and test the performances, which allows us to distinguish the two processes $g \rightarrow Q$-jet and $Q \rightarrow Q$-jet according to the final-state jet particle information. By isolating the $c \rightarrow c$-jet and $b \rightarrow b$-jet, the jets initiated by heavy quarks, we predict that the order of their $R_{AA}$ are in line with the mass hierarchy of energy loss. Measurements on the $R_{AA}$ of $Q \rightarrow Q$-jet and $g \rightarrow Q$-jet in heavy-ion collisions will provide a unique chance to test the flavor/mass dependence of energy loss at the jet level.

%%%%%%%%%%%%%%%%%%%%%%%%%%%%%%%%%%%%%%%%%%%%%%%%%%%%%%%%%%%%%%%%%%%%%
%%%%%%%%%%%%%%%%%%%%%%%%%%%%%%%%%%%%%%%%%%%%%%%%%%%%%%%%%%%%%%%%%%%%%
%%%%%%%%%%%%%%%%%%%%%%%%%%%%%%%%%%%%%%%%%%%%%%%%%%%%%%%%%%%%%%%%%%%%%
%%%%%%%%%%%%%%%%%%%%%%%%%%%%%%%%%%%%%%%%%%%%%%%%%%%%%%%%%%%%%%%%%%%%%
\vspace{0.0in}
\acknowledgments
This research is supported by the Guangdong Major Project of Basic and Applied Basic Research No. 2020B0301030008 and 2023A1515011460, and the National Natural Science Foundation of China Nos. 11935007, 12035007, 12247127 and 12375137. S. W. is supported by the Open Foundation of Key Laboratory of Quark and Lepton Physics (MOE) No. QLPL2023P01 and the Talent Scientific Star-up Foundation of the China Three Gorges University (CTGU) No. 2024RCKJ013.

\end{document}